\documentclass[showpacs]{revtex4}
\usepackage{amsmath}
\usepackage{amssymb}
\usepackage{graphicx}
\usepackage{color}
\usepackage{dcolumn}

\begin{document}
\newcolumntype{.}{D{.}{.}{-1}}

\title{
Magnetic fields in mixed neutron-star-plus-wormhole systems
}

\author{{\bf Ascar Aringazin$^{1}$}}
\email[{\it Email:}]{aringazin@gmail.com}
\author{{\bf Vladimir Dzhunushaliev$^{2,3,4,5}$}}
\email[{\it Email:}]{v.dzhunushaliev@gmail.com}
\author{{\bf Vladimir~Folomeev$^{2,4}$}}
\email[{\it Email:}]{vfolomeev@mail.ru}
\author{{\bf Burkhard Kleihaus $^2$}}
\email[{\it Email:}]{b.kleihaus@uni-oldenburg.de}
\author{{\bf Jutta Kunz$^2$}}
\email[{\it Email:}]{jutta.kunz@uni-oldenburg.de}
\affiliation{
$^1$ Institute for Basic Research,
Eurasian National University,
Astana 010008, Kazakhstan\\
$^2$
Institut f\"ur Physik, Universit\"at Oldenburg, Postfach 2503,
D-26111 Oldenburg, Germany\\
$^3$
Department of Theoretical and Nuclear Physics,
Al-Farabi Kazakh National University, Almaty 050040, Kazakhstan\\
$^4$Institute of Physicotechnical Problems and Material Science of the NAS
of the
Kyrgyz Republic, 265 a, Chui Street, Bishkek, 720071,  Kyrgyz Republic\\
$^5$ Institute of Experimental and Theoretical Physics,
Al-Farabi Kazakh National University, Almaty 050040, Kazakhstan
}

\begin{abstract}
We consider mixed configurations consisting of a wormhole filled
by a strongly magnetized isotropic or anisotropic neutron fluid.
The nontrivial topology of the spacetime is
allowed by the presence of exotic matter.
By comparing these configurations with ordinary magnetized neutron stars,
we clarify the question of how the presence of
the nontrivial topology influences the
magnetic field distribution inside the fluid.
In the case of an anisotropic fluid,
we find new solutions describing configurations,
where the maximum of the fluid density is shifted from the center.
A linear stability analysis shows that
these mixed configurations are unstable.
\end{abstract}

\date{\today}

\maketitle

\section{Introduction}

It is now widely believed that neutron stars possess strong magnetic fields.
These arise during the creation process of neutron stars,
when ordinary heavy stars with burnt-out cores
undergo catastrophic contraction after losing their hydrostatic stability.
In this case, since
the magnetic field is ``frozen'' into the stellar matter,
the magnetic flux through the surface of the star is conserved,
and the resultant compact objects have magnetic fields
with a strength of~$\sim 10^{8}-10^{13}~\text{gauss}$.
Even stronger magnetic fields on the order of $\sim 10^{15}~\text{gauss}$
are believed to exist in so-called
magnetars~\cite{Duncan:1992hi}.

Much research has been devoted to stellar magnetic fields,
including those existing in neutron stars
(see, e.g., the book~\cite{Mest} and more recent works
\cite{magn_stars,Kiuchi:2007pa} and references therein).
Studies show
that the structure of the interior magnetic fields and their stability
depend crucially on the properties of matter at extreme densities,
which are typical for the interior regions of a neutron star.
Such extreme magnetic fields play an important role
in the physics of neutron stars,
affecting their interior structure and determining their
evolution in time, such as the
magnetic dipole radiation, the deformation due to the magnetic stress, etc.

Here we show that, apart from the properties of the neutron matter,
the structure of the interior magnetic fields may also be
substantially influenced by the topology of the spacetime.
For this purpose,
we consider mixed configurations with nontrivial topology
which consist both of ordinary and exotic matter.
The latter is some special form of matter,
which implicates the violation of the null energy condition.
In the language of hydrodynamics this corresponds to
large negative effective pressures,
$p<-\varepsilon$, where $\varepsilon$ represents the energy density.

In our previous works
we have studied such mixed systems for the case
where the nontrivial (wormhole) topology is provided by exotic matter
in the form of a ghost scalar field,
and the ordinary matter is chosen in the form of relativistic neutron matter
(see Refs.~\cite{Dzhunushaliev:2011xx,Dzhunushaliev:2012ke,Dzhunushaliev:2013lna,Dzhunushaliev:2014mza},
where the motivation to consider this type of objects is also discussed).
The resulting neutron-star-plus-wormhole configurations
then possess properties of wormholes and of ordinary stars.
On the one hand, for a distant observer, they look like
ordinary neutron stars with typical masses and sizes.
On the other hand, such mixed systems may have
some new distinctive characteristics
which could, in principle, be traced in astrophysical observations
(see Refs.~\cite{Dzhunushaliev:2012ke,Dzhunushaliev:2014mza}).

Here we add to such mixed systems a magnetic field
associated with the neutron matter.
Our goal is to clarify the question of how the presence of
the wormhole influences the distribution and intensity of the magnetic field
inside the neutron matter.

To describe the structure of the magnetic field,
we employ a simplified model of a dipole magnetic field
as discussed in Ref.~\cite{Konno:1999zv}.
Namely, we consider
an axisymmetric, poloidal magnetic field
 produced by toroidal electric currents,
preserving the circularity property of spacetime~\cite{Bocquet:1995je}.
In the presence of such a field, the matter configurations
retain only axial symmetry.
However, since in ordinary neutron stars
the energy density of the magnetic field is
much smaller than the energy density of the neutron matter, we
%appearing in the system,
follow Ref.~\cite{Konno:1999zv} and
make a perturbative expansion of the field equations,
including perturbations up to second order.
We use these perturbed equations to derive an equation for the current
which cannot be arbitrarily chosen but follows from an integrability condition
for the perturbation equations.

%As observed before~\cite{Konno:1999zv,Sotani:2006at},
%deformations of a neutron star associated with
%neglect the small deformation of the neutron matter associated with
%the anisotropic pressure of the magnetic field are small
%for a field strength of~$\sim 10^{8}-10^{13}~\text{gauss}$.
%However, for anisotropic neutron matter the second order
%perturbations of the metric must be evaluated for consistency
%of the set of equations. For isotropic neutron matter, on the other hand,
%this is not necessary, and the problem simplifies accordingly.

The paper is organized as follows.
In Sec.~\ref{statem_prob} we present the statement of the problem
and derive the corresponding general-relativistic equations
for the mixed systems under consideration.
In Sec.~\ref{num_calc} we numerically solve these equations
for ordinary magnetized neutron stars
and for mixed configurations with different choices
for the parameters of the systems.
Comparing the results, we demonstrate the influence
of the nontrivial topology
on the structure and strength of the magnetic fields.
We perform a linear stability analysis
for these configurations in Sec.~\ref{stabi}.
Finally, in Sec.~\ref{conclusion} we summarize the results obtained.

\section{Statement of the problem}
\label{statem_prob}

Here we consider mixed systems
consisting of a gravitating ghost scalar field $\varphi$
(taken for simplicity to be massless)
and strongly magnetized neutron matter.
The scalar field allows for the presence of
a nontrivial wormhole-like topology of the system.
The wormhole is threaded by magnetized neutron matter.

The Lagrangian for this system  can be presented in the form
\begin{equation}
\label{lagran_wh_star_magn}
L=-\frac{c^4}{16\pi G_{\rm N}}R-\frac{1}{2}\partial_{\mu}\varphi\partial^{\mu}\varphi+L_{\text{fl}}-\frac{1}{4}F_{\mu\nu}F^{\mu\nu}~.
\end{equation}
Here $G_{\rm N}$ is the gravitational constant,
$F_{\mu\nu}$ is the electromagnetic field strength tensor,
and $L_{\text{fl}}$ is the Lagrangian of the neutron fluid.

\subsection{Neutron matter equation of state}
\label{fluid_EOS}

We assume that the neutron fluid can in general be
%is perfect and (possibly)
anisotropic.
The latter means that the radial and tangential components of its pressure
are not equal to each other.
Such a situation may arise at high densities
of the neutron matter~\cite{anis_reas}.
To describe the anisotropy, we here employ one of the approaches
of Refs.~\cite{Heintzmann:1975,Hillebrandt:1976},
and obtain the following components of the fluid energy-momentum tensor
(for details, see Ref.~\cite{Dzhunushaliev:2014mza}):
\begin{equation}
\label{fluid_emt_comp}
T_{t (\text{fl})}^t=\varepsilon, \quad T_{r (\text{fl})}^r=-p_{\rm rad},
\quad T_{\Theta (\text{fl})}^\Theta=T_{\phi (\text{fl})}^\phi=-p_{\rm tan},
\end{equation}
where the radial and the tangential components of the pressure are
\begin{equation}
\label{fluid_pressures}
p_{\rm rad}=(1-\alpha)p,  \quad p_{\rm tan}=(1+\alpha/2)p.
\end{equation}
Here $\alpha$ is a parameter determining the anisotropy
of the fluid, i.e., the anisotropy parameter,
and $p$ is the pressure appearing in the equation of state (EOS)
[see Eq.~\eqref{eqs_NS_WH} below].
Using these expressions, one can eliminate $p$,
and obtain the following relation between the pressure components:
\begin{equation}
\label{parameter_beta}
p_{\rm tan}=(1+\beta)p_{\rm rad} \quad \text{with} \quad
\beta=\frac{3}{2}\frac{\alpha}{1-\alpha}.
\end{equation}
%Thus $\beta$ (or equivalently $\alpha$) is the parameter determining the anisotropy
%of the fluid (i.e., the anisotropy parameter).

While at the moment the EOS of neutron matter at high densities
is still unknown,
the literature in the field offers numerous proposals for the EOS,
derived from a variety of physical assumptions~\cite{NM_EOS}.
For our purpose, we restrict ourselves
to a simplified variant of the EOS,
where a more or less realistic neutron matter EOS
is approximated in the form of a polytropic EOS.
This EOS can be taken in the following form:
\begin{equation}
\label{eqs_NS_WH}
p=K \rho_{b}^{1+1/n}, \quad \varepsilon = \rho_b c^2 +n p,
\end{equation}
with the constant $K=k c^2 (n_{b}^{(ch)} m_b)^{1-\gamma}$,
the polytropic index $n=1/(\gamma-1)$,
and $\rho_b=n_{b} m_b$ denotes the rest-mass density
of the neutron fluid.
Here $\varepsilon$ is the energy density of the fluid,
$n_{b}$ is the baryon number density,
$n_{b}^{(ch)}$ is a characteristic value of $n_{b}$,
$m_b$ is the baryon mass,
and $k$ and $\gamma$ are parameters
whose values depend on the properties of the neutron matter.

As in our previous investigations
concerning mixed star-plus-wormhole systems
\cite{Dzhunushaliev:2012ke,Dzhunushaliev:2013lna,Dzhunushaliev:2014mza},
we here, for simplicity, take only one set of parameters for the neutron fluid.
Namely, we choose
$m_b=1.66 \times 10^{-24}\, \text{g}$,
$n_{b}^{(ch)} = 0.1\, \text{fm}^{-3}$,
$k=0.1$, and $\gamma=2$~\cite{Salg1994}.
%These parameters correspond to a gas of baryons interacting
%via a vector-meson field, as described by Zel'dovich \cite{Zeld1961,Zeld}.
We employ these values for the parameters in the numerical calculations
of Sec.~\ref{num_calc}.

\subsection{Field equations}

Let us now consider the field equations for the
mixed neutron-star-plus-wormhole systems with a dipole magnetic field.
Strongly magnetized relativistic stars would, in general, possess
a nonspherical shape because of a substantial magnetic pressure.
However, for values of the magnetic field strength on the order of
$10^{12}-10^{13}~\text{gauss}$,
deviations from the spherical shape are negligible,
since the energy of the magnetic field is considerably smaller
than the gravitational energy~\cite{Sotani:2006at}, and
only for much higher values of the magnetic field strength
the deformation of the star will become substantial.
%~\cite{Sotani:2006at}.

Therefore, at first sight it seems appropriate to evaluate the metric
only to lowest order, i.e., in the spherically symmetric
approximation, where the deformation of the star is neglected.
Since we here neither consider the rotation of the star,
we may thus take the following static spherically symmetric
line element in polar Gaussian coordinates
\begin{equation}
\label{metric_wh_poten}
ds^2=e^{\nu}(dx^0)^2-dr^2-R^2 \left(d\Theta^2+\sin^2\Theta\, d\phi^2\right),
\end{equation}
where $\nu$ and $R$ are functions of the radial coordinate $r$ only,
and $x^0=c\, t$ is the time coordinate.
With this line element we then obtain a set of ordinary differential equations
for the metric functions, the fluid and the scalar field.
We refer to these equations as the {\sl background equations}.

Having solved the background equations,
and restricting to a dipole field,
we would like to evaluate the magnetic field in this
spherically symmetric background,
requiring a certain strength at the boundary of the star.
The Maxwell equation then involves a current as source term.
However, this current cannot be chosen arbitrarily.
It must satisfy an integrability condition
\cite{Bocquet:1995je,Konno:1999zv}.
In the case of isotropic neutron matter,
the current equation is easily integrated.
But in the case of anisotropic neutron matter
a consistent solution of the current equation
requires the determination of the second order perturbations of the metric.

In the following we first give the set of background equations.
Next we discuss the equation for the magnetic field
and the second order perturbations for the metric,
necessary for obtaining the integrability condition of the current.

\subsubsection{Background equations}

To obtain the background equations,
we use the metric \eqref{metric_wh_poten}
and take into account the energy-momentum tensor %\eqref{EMT_total},
(without an electromagnetic field)
\begin{equation}
\label{EMT_part}
T_{\nu }^\mu=(\varepsilon+p) u_\nu u^\mu -\left(\delta^\mu_\nu -\Delta_\nu^\mu\right) p
-\partial_{\nu} \varphi \partial^{\mu} \varphi+\frac{1}{2}\delta^\mu_\nu \partial_{i} \varphi \partial^{i} \varphi,
\end{equation}
where $\Delta_\nu^\mu=\{0, \alpha, -\alpha/2, -\alpha/2 \}$.

Then the Einstein equations can be written in the form
\begin{eqnarray}
\label{Einstein-00_poten}
&&-\left[2\frac{R^{\prime\prime}}{R}+\left(\frac{R^\prime}{R}\right)^2\right]+\frac{1}{R^2}
=\frac{8\pi G_{\rm N}}{c^4} T_t^t=\frac{8\pi G_{\rm N}}{c^4}\left(\varepsilon-\frac{1}{2}\varphi^{\prime 2}\right),
 \\
\label{Einstein-11_poten}
&&-\frac{R^\prime}{R}\left(\frac{R^\prime}{R}+\nu^\prime\right)+\frac{1}{R^2}
=\frac{8\pi G_{\rm N}}{c^4} T_r^r=\frac{8\pi G_{\rm N}}{c^4}\left[-(1-\alpha)p+\frac{1}{2}\varphi^{\prime 2}\right],
\\
\label{Einstein-22_poten}
&&\frac{R^{\prime\prime}}{R}+\frac{1}{2}\frac{R^\prime}{R}\nu^\prime+
\frac{1}{2}\nu^{\prime\prime}+\frac{1}{4}\nu^{\prime 2}
=-\frac{8\pi G_{\rm N}}{c^4} T_\Theta^\Theta=\frac{8\pi G_{\rm N}}{c^4}\left[\left(1+\frac{\alpha}{2}\right)p+\frac{1}{2}\varphi^{\prime 2}\right],
\end{eqnarray}
where the prime denotes differentiation with respect to $r$.

% with the components of the energy-momentum tensor \eqref{EMT_total}
% \begin{eqnarray}
%\label{emt_comp}
%&&T_0^0=\varepsilon_0-\frac{1}{2}\varphi_0^{\prime 2},\\
%&&T_1^1=-(1-\alpha)p_0+\frac{1}{2}\varphi_0^{\prime 2},\\
%&&T_2^2=T_3^3=-\left(1+\frac{\alpha}{2}\right)p_0-\frac{1}{2}\varphi_0^{\prime 2},
%\end{eqnarray}
%Here we have neglected the second-order contributions coming from the magnetic field.

Using these expressions, the $\mu=r$ component
of the conservation law of the total energy-momentum tensor,
$T^\nu_{\mu; \nu}=0$, yields
\begin{equation}
\label{conserv_2}
(1-\alpha)\frac{d p}{d r}+\frac{1}{2}\left[\varepsilon+(1-\alpha)p\,\right]\frac{d\nu}{d r}-3\alpha \frac{R^\prime}{R}p=0.
\end{equation}

%Introducing the new function $R$ defined by $e^{\mu}=R^2$ and

%where the prime denotes  differentiation with respect to $r$.
%Here the corresponding components of the scalar field energy-momentum tensor
%have been obtained by varying the Lagrangian \eqref{lagran_sf} with respect to the metric.

The field equation for the massless scalar field can be easily  integrated to give
\begin{equation}
\label{sf_poten}
\varphi^{\prime 2}=\frac{D^2}{R^4}e^{-\nu},
\end{equation}
where $D$ is an integration constant.

Thus we have four unknown functions~-- $R$, $\nu$, $p$, and $\varphi$~--
for which there are five equations,
\eqref{Einstein-00_poten}-\eqref{sf_poten},
only four of which are independent.

\subsubsection{Equation for the magnetic field}

Following~\cite{Konno:1999zv},
we choose an axisymmetric, poloidal magnetic field,
which is created by a 4-current
\begin{equation}
\label{current_eq}
j_{\mu}=(0,0,0,j_{\phi}) .
\end{equation}
For such a current
the electromagnetic 4-potential $A_{\mu}$
has only a $\phi$-component
\begin{equation}
\label{4_pot_eq}
A_{\mu}=(0,0,0,A_{\phi}) .
\end{equation}

Taking into account the nonvanishing components of
the electromagnetic field tensor
$F_{r \phi}=\partial A_{\phi}/\partial r$ and
$F_{\Theta \phi}=\partial A_{\phi}/\partial \Theta$,
the Maxwell equations yield
in the background metric \eqref{metric_wh_poten}
the following elliptic equation for $A_{\phi}$
\begin{equation}
\label{maxw_A}
\frac{\partial^2 A_{\phi}}{\partial r^2}+\frac{1}{2}\frac{\partial \nu}{\partial r}\frac{\partial A_{\phi}}{\partial r}+
\frac{1}{R^2}\frac{\partial^2 A_{\phi}}{\partial \Theta^2}-
\frac{1}{R^2}\cot \Theta\frac{\partial A_{\phi}}{\partial \Theta}=-\frac{1}{c} j_{\phi}.
\end{equation}

The solution of this equation is sought as an expansion of
the potential $A_{\phi}$
and the current $j_{\phi}$ as follows~\cite{Regge:1957,Konno:1999zv}:
\begin{eqnarray}
\label{expan_A}
&& A_{\phi}=\sum_{l=1}^\infty a_l(r)\sin \Theta \frac{d P_l(\cos \Theta)}{d\Theta},\\
&& j_{\phi}=\sum_{l=1}^\infty j_l(r)\sin \Theta \frac{d P_l(\cos \Theta)}{d\Theta},
\end{eqnarray}
where $P_l$ is the Legendre polynomial of degree $l$.
Substituting these expansions into Eq.~\eqref{maxw_A}, we obtain
\begin{equation}
\label{maxw_A_expan}
a_l^{\prime \prime}+\frac{1}{2}\nu^\prime a_l^\prime-\frac{l(l+1)}{R^2}a_l=-\frac{1}{c}j_l.
\end{equation}
The solution of this equation can be found
once the current~$j_l$ is specified.
%the equation for which will be derived in the next subsection.

In the present paper we only consider a dipole magnetic field,
i.e.,~$l=1$.
(For convenience, we hereafter drop the index 1
replacing $a_1$ and $j_1$ by $a$ and $j$, respectively.)
As discussed above, the current $j$ cannot be chosen arbitrarily,
since it must satisfy an integrability condition
%\cite{Bocquet:1995je,Konno:1999zv}
derived below.

\subsubsection{Integrability condition}

To obtain the integrability condition,
we make use of the fact
that the magnetic field induces only small deformations
in the shape of the geometry of the configuration.
To describe these small deviations from spherical symmetry,
we follow the approach adopted in Ref.~\cite{Konno:1999zv}
and expand
the metric in multipoles around the spherically symmetric spacetime.

The deformations of the metric, the fluid and the scalar field
of the configuration represent second order perturbations,
whereas the electromagnetic potential
and the current correspond to first order perturbations
\cite{Konno:1999zv}.
Then to second order the metric can be taken in the form
\begin{eqnarray}
\label{pert_metr}
 ds^2  &=&   e^{\nu (r)} \left\{ 1 + 2 \left[ h_{0}(r) + h_{2}(r)
          P_{2}( \cos \Theta ) \right] \right\} (dx^0)^2  %\nonumber \\
   -  \left\{ 1 + \frac{2 }{r}
          \left[ m_{0}(r) + m_{2}(r) P_{2}( \cos \Theta )
          \right] \right\} dr^2
          \nonumber \\
  & & -  R(r)^2 \left[ 1 + 2 k_{2}(r) P_{2}( \cos \Theta )
        \right] \left( d \Theta^2
        +  \sin^2 \Theta d \phi^2 \right) \!,
\end{eqnarray}
where $h_0$, $h_2$, $m_0$, $m_2$, and $k_2$ are the second order corrections
of the metric, and
$P_2$ denotes the Legendre polynomial of order~2.

Now we need to consider
the total energy-momentum tensor of the mixed system
with the electromagnetic field included
\begin{equation}
\label{EMT_total}
T_{\nu }^\mu=(\varepsilon+p) u_\nu u^\mu -\left(\delta^\mu_\nu -\Delta_\nu^\mu\right) p
-\partial_{\nu} \varphi \partial^{\mu} \varphi+\frac{1}{2}\delta^\mu_\nu \partial_{\alpha} \varphi \partial^{\alpha} \varphi
-F^\mu_\alpha F_\nu^\alpha+\frac{1}{4}\delta^\mu_\nu F_{\alpha \beta}F^{\alpha \beta}.
\end{equation}
%where $\Delta_\nu^\mu=\{0, \alpha, -\alpha/2, -\alpha/2 \}$.
Here the fluid energy density and pressure
and the scalar field are also expanded in second order
\begin{eqnarray}
\label{expans}
&&\varepsilon(r,\Theta)=\varepsilon_0+\frac{\varepsilon_0^\prime}{p_0^\prime}\left(\delta p_0+\delta p_2 P_2\right),\\
&&p(r,\Theta)=p_0+\delta p_0+\delta p_2 P_2,\\
&&\varphi(r,\Theta)=\varphi_0+\delta \varphi_0+\delta \varphi_2 P_2,
\end{eqnarray}
with the background solutions now denoted by $\varepsilon_0, p_0, \varphi_0$
and the perturbations
$\delta p_0, \delta p_2, \delta \varphi_0, \delta \varphi_2$.
Note that all perturbations depend on $r$ only.

Substituting the above expressions into the conservation law
of the energy-momentum tensor, $T^\mu_{\nu; \mu}=0$,
and using the metric \eqref{pert_metr},
we obtain the following $\nu=r$ and $\nu=\Theta$ components
\begin{eqnarray}
\label{p2expres}
&&(1-\alpha)\delta p_2^\prime=\left\{3\alpha \frac{R^\prime}{R}-\frac{(1-\alpha)p_0^\prime+\varepsilon_0^\prime}{(1-\alpha)p_0+\varepsilon_0}
\left[3\alpha \frac{R^\prime}{R}\frac{p_0}{p_0^\prime}-(1-\alpha)\right]
\right\}\delta p_2-\left[(1-\alpha)p_0+\varepsilon_0\right]h_2^\prime+3\alpha p_0 k_2^\prime-\frac{2}{3}\frac{a^\prime}{R^2}\frac{j}{c},\\
\label{p2expres2}
&&(2+\alpha)\delta p_2=-\left[(2+\alpha)p_0+2\varepsilon_0\right]h_2-3\alpha \frac{m_2 p_0}{r}-\frac{4}{3}\frac{a}{R^2}\frac{j}{c}.
\end{eqnarray}

It is clear that in order to find a consistent solution to these equations, the current
$j$ cannot be chosen arbitrarily but should be obtained from the condition of their joint integrability.
To do this, we differentiate Eq.~\eqref{p2expres2} and subtract the expression obtained from~\eqref{p2expres}.
The resulting equation for the current then is
[for the EOS~\eqref{eqs_NS_WH} with $\gamma=2$]:
%the integrability condition for these two equations gives
%we obtain the following equation for the current:
\begin{equation}
\label{ic_current}
j^\prime+F j+N=0
\end{equation}
 with
\begin{eqnarray}
\label{funF}
 F&=&\frac{1}{2(1-\alpha)\left[1+(2-\alpha)K\rho_b/c^2\right]}\Big\{
 -4\left(1+2 K\rho_b/c^2\right)\frac{R^\prime}{R}-2\left(1+4 K\rho_b/c^2\right)\frac{\rho_b^\prime}{\rho_b}\nonumber\\
&+& 4\alpha^2\left(\frac{3}{4}\frac{a^\prime}{a}-\frac{R^\prime}{R}-\frac{\rho_b^\prime}{\rho_b}\right) K\frac{\rho_b}{c^2}+
\alpha\left[-3\left(1+2 K\rho_b/c^2\right)\frac{a^\prime}{a}+\left(1+12 K\rho_b/c^2\right)\frac{R^\prime}{R}+2\left(1+6 K\rho_b/c^2\right)\frac{\rho_b^\prime}{\rho_b}\right]
 \Big\}
\end{eqnarray}
and
\begin{eqnarray}
\label{funN}
N&=& \alpha \frac{9 c^3 R^2 \rho_b}{4(1-\alpha)a}\Big\{
-\left(1+K\rho_b/c^2\right)h_2^\prime+(2+\alpha)\frac{K\rho_b}{c^2} k_2^\prime+(1-\alpha)\frac{K\rho_b}{c^2} \frac{m_2^\prime}{r}\nonumber\\
&-&\frac{3\alpha r R^\prime/R+2(1-\alpha)\left[1+(2-\alpha)K\rho_b/c^2-r \rho_b^\prime/\rho_b\right]}{2\left[1+(2-\alpha)K\rho_b/c^2\right]}\frac{K\rho_b}{c^2}\frac{m_2}{r^2}\nonumber\\
&-&\frac{\left[2+(4+\alpha)K\rho_b/c^2\right]R^\prime/R-2(1-\alpha)K\rho_b^\prime/c^2}{2\left[1+(2-\alpha)K\rho_b/c^2\right]}h_2
\Big\}.
\end{eqnarray}

We note that all terms in Eq.~\eqref{ic_current} should be of first order.
Since $j$ itself is of first order,
the function $F$ must be of zeroth order.
On the other hand, the function $N$ must be of first order.
%Here we note, that
Since the potential $a$,
which is of first order, enters in the denominator,
we must keep terms up to second order
in the numerator for consistency.

Now we must consider a distinction of cases:
\begin{itemize}
\item {\sl isotropic case}: $\alpha=0$
\item[]
Here $N=0$, and Eq.~\eqref{ic_current} can be integrated to give
\begin{equation}
\label{ic_current_isot}
j=c_0 R^2 \rho_b\left(1+2 K\rho_b/c^2\right),
\end{equation}
where $c_0$ is an integration constant.
This expression corresponds to the one obtained in Ref.~\cite{Konno:1999zv}
[see their Eq.~(25)], written for our EOS~\eqref{eqs_NS_WH}.
\item[] Here the second order perturbations of the metric
and the fluid as well as the scalar field are not needed
to obtain a consistent solution in first order.
\item {\sl anisotropic case}: $\alpha \ne 0$
\item[]
Now $N\ne 0$, and a consistent treatment in first order
needs the second order perturbations.
The functions $h_2, k_2$, and $m_2$ appearing in \eqref{funN}
are determined by the following set of equations
deduced from the Einstein equations
\begin{eqnarray}
\label{pert_Einst_G11_l2}
&&h_2^\prime+\left(1+\frac{1}{2}\nu^\prime\frac{R}{R^\prime}\right)k_2^\prime=
\left(\nu^\prime+\frac{R^\prime}{R}\right)\frac{m_2}{r}+\frac{2}{R R^\prime}k_2+\frac{3}{R R^\prime}h_2 \nonumber \\
&&+\frac{4\pi G_{\rm N}}{c^4}\left\{
(1-\alpha)\delta p_2-\frac{1}{3 R^2}\left(a^{\prime 2}+\frac{4}{R^2}a^2\right)-\varphi_0^\prime \delta\varphi_2^\prime+\varphi_0^{\prime 2}\frac{m_2}{r}
\right\}\frac{R}{R^\prime},\\
\label{pert_Einst_G21_l2}
&&h_2^\prime+k_2^\prime=\left(\frac{R^\prime}{R}-\frac{1}{2}\nu^\prime\right)h_2+\left(\frac{R^\prime}{R}+\frac{1}{2}\nu^\prime\right)\frac{m_2}{r}+
\frac{8\pi G_{\rm N}}{c^4}\left(\frac{2}{3}\frac{a a^\prime}{R^2}+\varphi_0^\prime \delta \varphi_2\right), \\
\label{pert_Einst_G3_G22}
&& h_2+\frac{m_2}{r}=\frac{8\pi G_{\rm N}}{c^4}\frac{a^{\prime 2}}{3},\\
\label{pert_Einst_G00_l2}
&& k_2^{\prime\prime}+3\frac{R^\prime}{R}k_2^\prime-\left[2\frac{R^{\prime\prime}}{R}+\left(\frac{R^{\prime}}{R}\right)^2+\frac{3}{R^2}\right]\frac{m_2}{r}-
\frac{2}{R^2}k_2-\frac{R^\prime}{R}\left(\frac{m_2}{r}\right)^\prime \nonumber \\
&& =\frac{4\pi G_{\rm N}}{c^4}\left(-\frac{\varepsilon_0^\prime}{p_0^\prime}\delta p_2+\varphi_0^\prime\delta\varphi_2^\prime-\varphi_0^{\prime 2}\frac{m_2}{r}-
\frac{4}{3}\frac{a^2}{R^4}+\frac{1}{3}\frac{a^{\prime 2}}{R^2}\right).
\end{eqnarray}
%Here $G_{\rm N}$ is the gravitational constant.
In turn, the scalar field equation gives the perturbation equation
\begin{equation}
\label{pert_sfe}
\delta \varphi_2^{\prime \prime}+\left(\frac{1}{2}\nu^\prime+2\frac{R^\prime}{R}\right)\delta \varphi_2^{\prime}-\frac{6}{R^2}\delta \varphi_2+
\left[h_2^\prime+2 k_2^\prime-\left(\frac{m_2}{r}\right)^\prime\right]\varphi_0^\prime=0.
\end{equation}
These equations are then solved, in order to determine the
current in a consistent way.
\end{itemize}

\subsubsection{Scheme}

To obtain the solutions in first order, we must therefore
consider the isotropic and anisotropic cases separately.
\begin{itemize}
\item
In the isotropic case
we have five unknown functions~--
$R$, $\nu$, $p_0$, $\varphi_0$, and $a$~--
for which there are six equations,
\eqref{Einstein-00_poten}-\eqref{sf_poten},
\eqref{maxw_A_expan},
only five of which are independent.
These equations are supplemented by the EOS
\eqref{eqs_NS_WH}, and also by the expression for the current $j$,
Eq.~\eqref{ic_current_isot}. % (for the isotropic fluid).
\item
In the anisotropic case we must solve in addition
Eq.~\eqref{ic_current} %  (for the anisotropic fluid) or
together with the equations for the second order perturbations,
Eqs.~\eqref{pert_Einst_G11_l2}-\eqref{pert_sfe} and
Eq.~\eqref{p2expres2}.
\end{itemize}

%the latter can be taken in the form
%\begin{equation}
%\label{current_gen}
%j_1(r)=c_0 r^2(\varepsilon+p),
%\end{equation}
%where $c_0$ is an arbitrary constant. Such a current is used in
%Refs.~\cite{Konno:1999zv,Sotani:2006at} in modeling the magnetic field of a neutron star. This will allow us below to compare distributions of the
%magnetic field in our mixed systems and ordinary neutron stars obtained for the
%fixed form of the
%current given by Eq.~\eqref{current_gen}.

\section{Numerical results}
\label{num_calc}

We now discuss our results, obtained by numerically integrating
the equations for magnetized neutron stars and for
magnetized mixed systems with isotropic ($\beta=0$)
as well as anisotropic ($\beta\neq 0$) fluids.

\subsection{Neutron stars with magnetic field}
\label{NSmagn}

Since we want to compare the magnetic field distributions
of mixed systems with a nontrivial spacetime topology
with those of ordinary neutron stars,
we begin with the simple case of magnetized
ordinary neutron stars, where no scalar field is present
and the neutron fluid is isotropic.

We follow Ref.~\cite{Konno:1999zv},
but employ the EOS~\eqref{eqs_NS_WH} for the neutron matter
(instead of the one used in~\cite{Konno:1999zv}).
Working in first order, and considering the isotropic case,
there is no need to evaluate second order contributions
to the metric and the fluid.

It is convenient to rewrite the general set of equations
\eqref{Einstein-00_poten}-\eqref{conserv_2}
and \eqref{maxw_A_expan}
in terms of the following dimensionless variables
\begin{equation}
\label{dimless_var_NS}
\xi=\frac{r}{L}, \quad \Sigma=\frac{R}{L},
\quad \bar{a}(\xi)=\frac{8\pi G_{\rm N} \sqrt{\rho_{b c}}}{c^3}\,a(r),
\quad \text{where} \quad L=\frac{c}{\sqrt{8\pi G_{\rm N} \rho_{b c}}}.
\end{equation}
Moreover, we parametrize the fluid density as follows~\cite{Zeld}:
\begin{equation}
\label{theta_def}
\rho_b=\rho_{b c}\, \theta^n,
\end{equation}
where $\rho_{b c}$ is the rest-mass density of the neutron fluid at the
center of the star.

The dimensionless equations then take the form
\begin{eqnarray}
\label{Einstein-00_dmls_NS}
&&-\left[2\frac{\Sigma^{\prime\prime}}{\Sigma}+\left(\frac{\Sigma^\prime}{\Sigma}\right)^2\right]+\frac{1}{\Sigma^2}
=  (1+\sigma n \theta) \theta^n,
 \\
\label{Einstein-11_dmls_NS}
&&-\frac{\Sigma^\prime}{\Sigma}\left(\frac{\Sigma^\prime}{\Sigma}+\nu^\prime\right)+\frac{1}{\Sigma^2}
=-  \sigma \theta^{n+1},
\\
\label{Einstein-22_dmls_NS}
&&\frac{\Sigma^{\prime\prime}}{\Sigma}+\frac{1}{2}\frac{\Sigma^\prime}{\Sigma}\nu^\prime+
\frac{1}{2}\nu^{\prime\prime}+\frac{1}{4}\nu^{\prime 2}
=
 \sigma  \theta^{n+1},
\\
\label{fluid_dmls_NS}
&&\sigma(n+1)\theta^\prime+\frac{1}{2}\left[1+\sigma(n+1)\theta\right]\nu^\prime=0,\\
\label{Maxw_dmls_NS}
&&\bar{a}^{\prime \prime}+\frac{1}{2}\nu^\prime \bar{a}^\prime-\frac{2}{\Sigma^2}\bar{a}=-\bar{j}\equiv
-A_0\, \Sigma^2\left(1+2\sigma\theta\right)\theta.
\end{eqnarray}
Here $\sigma=K \rho_{b c}^{1/n}/c^2=p_c/(\rho_{b c} c^2)$ is a constant
related to the pressure $p_c$ of the fluid at the center.
The values of the fluid parameters appearing here
are taken from the end of Sec.~\ref{fluid_EOS}
(including the value $\gamma=2$ corresponding to $n=1$).
The expression for the dimensionless current
$\bar{j}=j/(c^2\sqrt{\rho_{b c}})$,
used in Eq.~\eqref{Maxw_dmls_NS}, is derived from \eqref{ic_current_isot},
and $A_0=c_0/(8\pi G_{\rm N} \sqrt{\rho_{b c}})$
[with $c_0$ the integration constant of $j$,
Eq.~(\ref{ic_current_isot})]
is an arbitrary dimensionless constant.

The boundary conditions at the center are given by
\begin{equation}
\label{bound_cond_NS_magn}
\Sigma\approx \xi+\frac{1}{6}\, \Sigma_3 \xi^3, \quad \nu \approx \nu_c+\frac{1}{2}\, \nu_2 \xi^2, \quad
\theta\approx \theta_c+\frac{1}{2}\, \theta_2 \xi^2, \quad \bar{a} \approx a_c\, \xi^2,
\end{equation}
where $\nu_c$  and $\theta_c=1$ are
the central values of the corresponding functions,
and the coefficients $\Sigma_3, \nu_2$, and $\theta_2$ are
%appearing in  \eqref{bound_cond_NS_magn}
\begin{equation}
\label{bound_coef_NS_magn}
\Sigma_3=-\frac{1}{3}\left(1+\sigma n \theta_c\right)\theta_c^n, \quad
\nu_2=\sigma \theta_c^{n+1}-\Sigma_3, \quad
\theta_2=-\frac{1+\sigma(n+1)\theta_c}{2\sigma(n+1)}\nu_2.
\end{equation}
%The constant $a_c$ is fixed
%by the boundary condition at the surface of the neutron star,
%chosen to achieve a prescribed value of the surface magnetic field.

Using these boundary conditions, we numerically solved
the set of Eqs.~\eqref{Einstein-00_dmls_NS}-\eqref{Maxw_dmls_NS}.
Starting the integration near the origin (i.e., $\xi \approx 0$)
we integrated outwards up to the point $\xi=\xi_b$,
where the function $\theta$ became zero,
corresponding to the boundary of the fluid.
(Note that the boundary of the fluid $\xi_b$ is defined by $p(\xi_b)=0$.)

At the boundary of the fluid
these interior solutions were then matched
with the exterior solutions
by equating the corresponding values of the functions
$\Sigma$, $\nu$, $\bar{a}$ and their derivatives.
The exterior solutions in turn were obtained
by requiring asymptotic flatness.
Knowledge of the asymptotic solutions then allowed
to determine the value of the integration constant $\nu_c$ at the center.

The free parameters $a_c$ and $A_0$ of the electromagnetic potential
were chosen such that the magnetic field \eqref{streng_magn_dmls_NS}
%at the surface of the star
has a surface value
typical for magnetic neutron stars, i.e., $\sim 10^{12}~\text{gauss}$,
while asymptotically the field is vanishing.

In order to get a better understanding of the magnetic field,
we consider its tetrad components
\begin{equation}
\label{streng_magn_dmls_NS}
B_{\hat{r}}=-F_{\hat{\Theta} \hat{\phi}}=2 c\sqrt{\rho_{b c}}\frac{\bar{a}}{\Sigma^2}\cos{\Theta}, \quad
B_{\hat{\Theta}}=F_{\hat{r} \hat{\phi}}=- c\sqrt{\rho_{b c}}\frac{\bar{a}^\prime}{\Sigma}\sin{\Theta}.
\end{equation}
Fig.~\ref{MF_fig} exhibits these tetrad components together with
the current distribution for a neutron star
with central rest-mass density
$\rho_{b c}= 5.37 \times 10^{14} \text{g cm}^{-3}$.
%$L=10\, \text{km}$, and $\beta=0$ (isotropic fluid).
The component $B_{\hat{r}}$ is calculated on the symmetry axis ($\Theta=0$),
and the component  $B_{\hat{\Theta}}$ --
in the equatorial plane ($\Theta=\pi/2$).

We note that an ordinary magnetized (isotropic)
%neutron star has finite values
%of $B_{\hat{r}}$ ($\Theta=0$)
%and $B_{\hat{\Theta}}$ ($\Theta=\pi/2$) at the center.
neutron star has a nonvanishing
magnetic field at the center.
In contrast, its current vanishes at the center.
The corresponding magnetic field lines are exhibited
in Fig.~\ref{fig_magn_field_lines}.
Finally, the corresponding energy density
$T_{0 (\text{fl})}^0=\varepsilon$ of the neutron fluid
[see Eqs.~\eqref{fluid_emt_comp} and \eqref{eqs_NS_WH}]
is shown in Fig.~\ref{fig_NF_energ_dens}.

\subsection{Mixed neutron-star-plus-wormhole systems}
\label{Mixedmagn}

\subsubsection{Background equations}

Following Ref.~\cite{Dzhunushaliev:2014mza},
we rewrite the equations \eqref{Einstein-00_poten}-\eqref{sf_poten}
in terms of another set of dimensionless variables.
For that purpose we consider the massless scalar field $\varphi$.
Without loss of generality,
its value at the center of the configuration, i.e., at $r=0$,
can be taken equal to zero,
while its derivative at $r=0$ is nonzero.
In the neighborhood of the center
the scalar field can be expanded as
\begin{equation}
\varphi_0 \approx \varphi_1 r +\frac{1}{6}\varphi_3 r^3,
\end{equation}
where $\varphi_1$ is the derivative at the center.
Its square is a measure for the ``kinetic'' energy of the scalar field.

It is now convenient to employ the following dimensionless variables
expressed in terms of $\varphi_1$:
%In particular, we introduce
\begin{equation}
\label{dimless_xi_v}
\xi=\frac{r}{L}, \quad \Sigma=\frac{R}{L},
\quad \phi_0(\xi)=\frac{\sqrt{8\pi G_{\rm N}}}{c^2}\,\varphi_0(r),
%\quad \bar{a}(\xi)=\frac{8\pi G_{\rm N} \varphi_1}{c^4}\,a(r), \quad \bar{j}(\xi)=\frac{j(r)}{c\varphi_1},
\quad \text{where} \quad L=\frac{c^2}{\sqrt{8\pi G_{\rm N}}\varphi_1}.
\end{equation}
We then rewrite %Eqs.~\eqref{maxw_A_expan},  \eqref{ic_current},
Eqs.~\eqref{Einstein-00_poten}-\eqref{sf_poten} in the dimensionless form
\begin{eqnarray}
\label{Einstein-00_dmls}
&&-\left[2\frac{\Sigma^{\prime\prime}}{\Sigma}+\left(\frac{\Sigma^\prime}{\Sigma}\right)^2\right]+\frac{1}{\Sigma^2}
={\cal B}  (1+\sigma n \theta) \theta^n-\frac{1}{2}\phi_0^{\prime 2},
 \\
\label{Einstein-11_dmls}
&&-\frac{\Sigma^\prime}{\Sigma}\left(\frac{\Sigma^\prime}{\Sigma}+\nu^\prime\right)+\frac{1}{\Sigma^2}
=-{\cal B}  \sigma (1-\alpha) \theta^{n+1}+\frac{1}{2}\phi_0^{\prime 2},
\\
\label{Einstein-22_dmls}
&&\frac{\Sigma^{\prime\prime}}{\Sigma}+\frac{1}{2}\frac{\Sigma^\prime}{\Sigma}\nu^\prime+
\frac{1}{2}\nu^{\prime\prime}+\frac{1}{4}\nu^{\prime 2}
=
{\cal B} \sigma \left(1+\frac{\alpha}{2}\right) \theta^{n+1}
+\frac{1}{2}\phi_0^{\prime 2},
\\
%\label{sf_dmls}
%&&\phi_0^{\prime 2}=\frac{e^{\nu_c-\nu}}{(\Sigma/\Sigma_c)^4},
%\\
\label{fluid_dmls}
&&\sigma(n+1)(1-\alpha)\theta^\prime+\frac{1}{2}\left[1+\sigma(n+1-\alpha)\theta\right]\nu^\prime
-3\alpha \sigma\theta\frac{\Sigma^\prime}{\Sigma}=0,\\
\label{sf_dmls}
&&\phi_0^{\prime 2}=\frac{e^{\nu_c-\nu}}{(\Sigma/\Sigma_c)^4}.
%\label{Maxw_dmls}
%&&\bar{a}^{\prime \prime}+\frac{1}{2}\nu^\prime \bar{a}^\prime-\frac{2}{\Sigma^2}\bar{a}=-\bar{j}, \\
%\label{current_anis_dmls}
%&&\bar{j}^\prime+\bar{F} \bar{j}+\bar{N}=0.
\end{eqnarray}
Here
${\cal B}=(\rho_{b c} c^2)/\varphi_1^2$ is a dimensionless quantity,
providing a measure of the ratio of the rest-energy density
of the fluid to the energy density of the scalar field at the center;
$\Sigma_c$ and $\nu_c$ are the central values of the corresponding functions
[see Eq.~\eqref{bound_cond_all} below];
and the integration constant $D$ from Eq.~\eqref{sf_poten}
is chosen as $D^2=\left(c^4/8\pi G_{\rm N} \varphi_1\right)^2 \Sigma_c^4 e^{\nu_c}$
to provide $\phi_0^\prime=1$ at the center.

Let us now address the boundary conditions for the mixed systems.
We consider asymptotically flat configurations, that are symmetric under
$\xi \to -\xi$.
As discussed in \cite{Dzhunushaliev:2014mza},
the metric function $\Sigma(\xi)$ can possess either a minimum
at $\xi=0$,
corresponding to a single-throat configuration, or it can
possess a local maximum,
corresponding to a double-throat configuration.
In any case, in the neighborhood of the center
the following expansion is appropriate
\begin{equation}
\label{bound_cond_all}
\Sigma\approx \Sigma_c+\frac{1}{2}\, \Sigma_2 \xi^2, \quad \nu \approx \nu_c+\frac{1}{2}\, \nu_2 \xi^2, \quad
\theta\approx \theta_c+\frac{1}{2}\, \theta_2 \xi^2, %\quad \bar{a} \approx a_c+1/2\, a_2 \xi^2,
%\quad \bar{j}=j_c+1/2 \, j_2 \xi^2.
\end{equation}
%Notice that unlike the neutron star, which has $\bar{a}, \bar{j}=0$ at the center [see Eq.~\eqref{bound_cond_NS_magn}],
%here we must take the nonzero central values of these variables
%%$\bar{a}=a_c$ at the center
%to obtain asymptotically flat solutions.
%Using these conditions in Eqs.~\eqref{Einstein-00_dmls}-\eqref{Maxw_dmls}, we find
where $\theta_c$ is chosen to be 1 in subsequent numerical calculations.
Inserting this expansion in Eqs.~\eqref{Einstein-00_dmls}-\eqref{sf_dmls},
we find the coefficients
\begin{eqnarray}
\label{bound_coef}
&&\Sigma_c=\frac{1}{\sqrt{1/2-{\cal B} \sigma(1-\alpha)\theta_c^{n+1}}}, \quad
\Sigma_2=\frac{\Sigma_c}{2}\left[\frac{1}{2}+\frac{1}{\Sigma_c^2}-{\cal B}\left(1+\sigma n\theta_c\right)\theta_c^n\right], \nonumber\\
&&\nu_2=2\left[\frac{1}{2}+{\cal B}\sigma\left(1+\frac{\alpha}{2}\right)\theta_c^{n+1}-\frac{\Sigma_2}{\Sigma_c}\right],\\
&&\theta_2=\frac{1}{\sigma(n+1)(1-\alpha)}\Big\{3\alpha \sigma \theta_c \frac{\Sigma_2}{\Sigma_c}-
\frac{1}{2}\left[1+\sigma(n+1-\alpha)\theta_c\right]\nu_2\Big\}.
%\quad a_2=2\frac{a_c}{\Sigma_c^2}-j_c.
\nonumber
\end{eqnarray}
%with $g=1/2-B\sigma(1-\alpha)\theta_c^{n+1}$.
%Then, complementing
The conditions \eqref{bound_cond_all}
%\eqref{bound_cond_all} must be
are supplemented by the conditions for the scalar field
at the center, $\phi_0(0) =0,\  \phi_0^\prime (0)=1$.

The asymptotic behavior of the background solutions is given by
\begin{equation}
\label{asym_ci}
 \phi_0 \to C_1 -\frac{C_2}{\xi}, \quad
\Sigma \to \xi, \quad \Sigma^\prime \to 1-\frac{C_3}{\xi},
 \quad e^\nu \to 1-2 \frac{C_3}{\xi}~,
\end{equation}
where the $C_i$ are integration constants.

With all this at hand,
Eqs.~\eqref{Einstein-00_dmls}-\eqref{sf_dmls} can be solved numerically.
But before discussing the results,
let us address two features of the mixed systems.

First, we note that depending on the parameters of the system
the coefficient $\Sigma_2$ can be either positive or negative
[see Eq.~\eqref{bound_coef}]. This determines
whether the mixed configurations possess a single throat
at the center or an equator surrounded by a double throat
(cf. Ref.~\cite{Dzhunushaliev:2014mza}).

We previously showed~\cite{Dzhunushaliev:2014mza}
that the behavior of the metric function $\Sigma(\xi)$
in the neighborhood of the center of the system,
$\xi=0$, depends strongly on the value of the parameter ${\cal B}$.
For small values of ${\cal B}$,  there is a single throat,
located at the center of the configuration.
But for increasing values of ${\cal B}$
the center of the configuration no longer represents a throat
but instead corresponds to an equator.
On each side of the equator a minimal area surface
and thus a throat is located.
The resulting configurations represent double-throat
systems, where the throats can be or not be filled by the fluid.
The expression for ${\cal B}$,
\begin{equation}
\label{B_throu_L}
{\cal B}=8\pi G_{\rm N} \rho_{b c} (L/c)^2,
\end{equation}
shows that the value of ${\cal B}$ is determined
by fixing $L$ and $\rho_{b c}$.

Second, the presence of an anisotropic pressure of the neutron fluid
permits us to obtain solutions describing configurations
with a maximum value of the density of the fluid $\rho_{\text{max}}$
that is not located at the center but at some $\xi\neq 0$.
Indeed, as seen from the expression for
$\theta_2$ in Eq.~\eqref{bound_coef},
in the presence of anisotropy,
i.e., for $\alpha\neq 0$,
$\theta_2$ can become positive.
Thus the function $\theta$ can possess
a local minimum at the center.

This situation is similar to the one encountered for
ordinary neutron stars
supported by an anisotropic fluid~\cite{Heintzmann:1975, Hillebrandt:1976}.
In this case the central density of the neutron fluid $\rho_{b c}$
represents a characteristic parameter,
which affects the value of the parameter
${\cal B}$ as well [see Eq.~\eqref{B_throu_L}].
As shown in Ref.~\cite{Dzhunushaliev:2014mza},
this parameter in effect determines the structure of the resulting solutions
and physical parameters (masses and sizes)
of the configurations under consideration.

In this case there is some value ${\cal B}_{p}$
at which $\theta_2$ passes through~0 and becomes negative.
This then returns us to ordinary configurations
where the maximum density of the neutron fluid resides at the center.
Only for $0<{\cal B}<{\cal B}_{p}$
$\theta_2$ is positive, lying within the bounds
\begin{equation}
\frac{3\alpha \theta_c}{2(n+1)(1-\alpha)}\gtrsim\theta_2>0
\end{equation}
(where the left restriction is obtained in the limit ${\cal B}\rightarrow 0$).
Note that we here consider only positive values of $\alpha$,
for which the above inequality is valid.
%It is necessary since, in principle, $\alpha$ may be negative, and then (55) should be modified.
As the anisotropy increases, i.e., as $\beta$ increases,
$\alpha \rightarrow 1$ from below,
resulting in an increase of $\theta_2$
and correspondingly an increase of the maximum density $\rho_{\text{max}}$.
%As a result,
This leads to changes in the masses and sizes
of the resulting configurations, as discussed below.

\subsubsection{Magnetic field equations}

We now turn to the set of perturbative equations
involving the determination of the electromagnetic potential
and the current.
Analogous to the background equations,
we here introduce appropriate dimensionless variables
\begin{equation}
\label{dimless_xi_v2}
\quad \bar{a}(\xi)=\frac{8\pi G_{\rm N} \varphi_1}{c^4}\,a(r),
\quad \bar{j}(\xi)=\frac{j(r)}{c\varphi_1} .
\end{equation}
Then we rewrite Eqs.~\eqref{maxw_A_expan} and \eqref{ic_current}
in dimensionless form
\begin{eqnarray}
\label{Maxw_dmls}
&&\bar{a}^{\prime \prime}+\frac{1}{2}\nu^\prime \bar{a}^\prime-\frac{2}{\Sigma^2}\bar{a}=-\bar{j}, \\
\label{current_anis_dmls}
&&\bar{j}^\prime+\bar{F} \bar{j}+\bar{N}=0,
\end{eqnarray}
where
$\bar{F}, \bar{N}$ are the dimensionless expressions
obtained from $F$ and $N$, Eqs.~\eqref{funF} and \eqref{funN}.

To determine the boundary conditions,
we again make an expansion in the neighborhood of the center
\begin{equation}
\label{bound_cond_all2}
\bar{a} \approx a_c+\frac{1}{2}\, a_2 \xi^2,
\quad \bar{j} \approx j_c+\frac{1}{2} \, j_2 \xi^2.
\end{equation}
Note that unlike for the neutron star,
which has $\bar{a}, \bar{j}=0$ at the center
[see Eq.~\eqref{bound_cond_NS_magn}],
we here must allow for nonzero central values of these variables
at $\xi=0$,
%$\bar{a}=a_c$ at the center
to obtain asymptotically flat solutions.

Inserting these conditions
in Eq.~\eqref{Maxw_dmls}, we find %-\eqref{current_anis_dmls}, we find
\begin{eqnarray}
\label{bound_coef2}
\quad a_2=2\frac{a_c}{\Sigma_c^2}-j_c.\nonumber
\end{eqnarray}

\begin{itemize}
\item
In the isotropic case $\bar N=0$,
and Eq.~\eqref{current_anis_dmls} can be solved to give
\begin{equation}
\label{current_mix_iso}
\bar{j}=A_1 {\cal B}\, \Sigma^2\left[1+2\sigma\theta\right]\theta,
\end{equation}
where $A_1=c_0 c/(8\pi G_{\rm N} \varphi_1)$ is an arbitrary dimensionless constant.
Subsequently Eq.~(\ref{Maxw_dmls}) can be solved.
\item
When an anisotropy is present,
together with Eqs.~(\ref{Maxw_dmls}) and~\eqref{current_anis_dmls}
the Einstein equations~\eqref{pert_Einst_G11_l2}-\eqref{pert_Einst_G00_l2}
and the scalar-field equation~\eqref{pert_sfe}
for the second order perturbations need to be solved.
Here we choose the boundary conditions for them in the form
\begin{equation}
\label{bound_cond_pert}
h_2 \approx h_{2c}+\frac{1}{2}h_{22}\xi^2,
\quad k_2 \approx k_{2c}+\frac{1}{2}k_{22}\xi^2,
\quad \delta \phi_2 \approx \delta \phi_{21}\xi.
\end{equation}
The parameters $a_c, j_c, h_{2c}, k_{2c}, \delta \phi_{21}$
appearing in \eqref{bound_cond_all2} and \eqref{bound_cond_pert}
are chosen in such a way as to provide
regularity of the perturbed solutions at the center,
$\xi=0$, and to obtain asymptotically decaying solutions for
$\xi \rightarrow \infty$.
Taking into account the asymptotic behavior of the background solutions,
Eq.~(\ref{asym_ci}),
we find for the second order functions the asymptotic behavior
$
\delta \phi_2 \sim \xi^{-3}, \; h_2, k_2 \sim \xi^{-4}.
$
\end{itemize}

Finally, we take the tetrad components of the magnetic field
in the form
\begin{equation}
\label{streng_magn_dmls_mixed}
B_{\hat{r}}=2 c\sqrt{\frac{\rho_{b c}}{{\cal B}}}\frac{\bar{a}}{\Sigma^2}\cos{\Theta}, \quad
B_{\hat{\Theta}}=- c\sqrt{\frac{\rho_{b c}}{{\cal B}}}\frac{\bar{a}^\prime}{\Sigma}\sin{\Theta}.
\end{equation}

\subsubsection{Results for isotropic and anisotropic fluids}

\begin{figure}[p!]
\begin{minipage}{1\linewidth}
  \begin{center}
  \includegraphics[width=17cm]{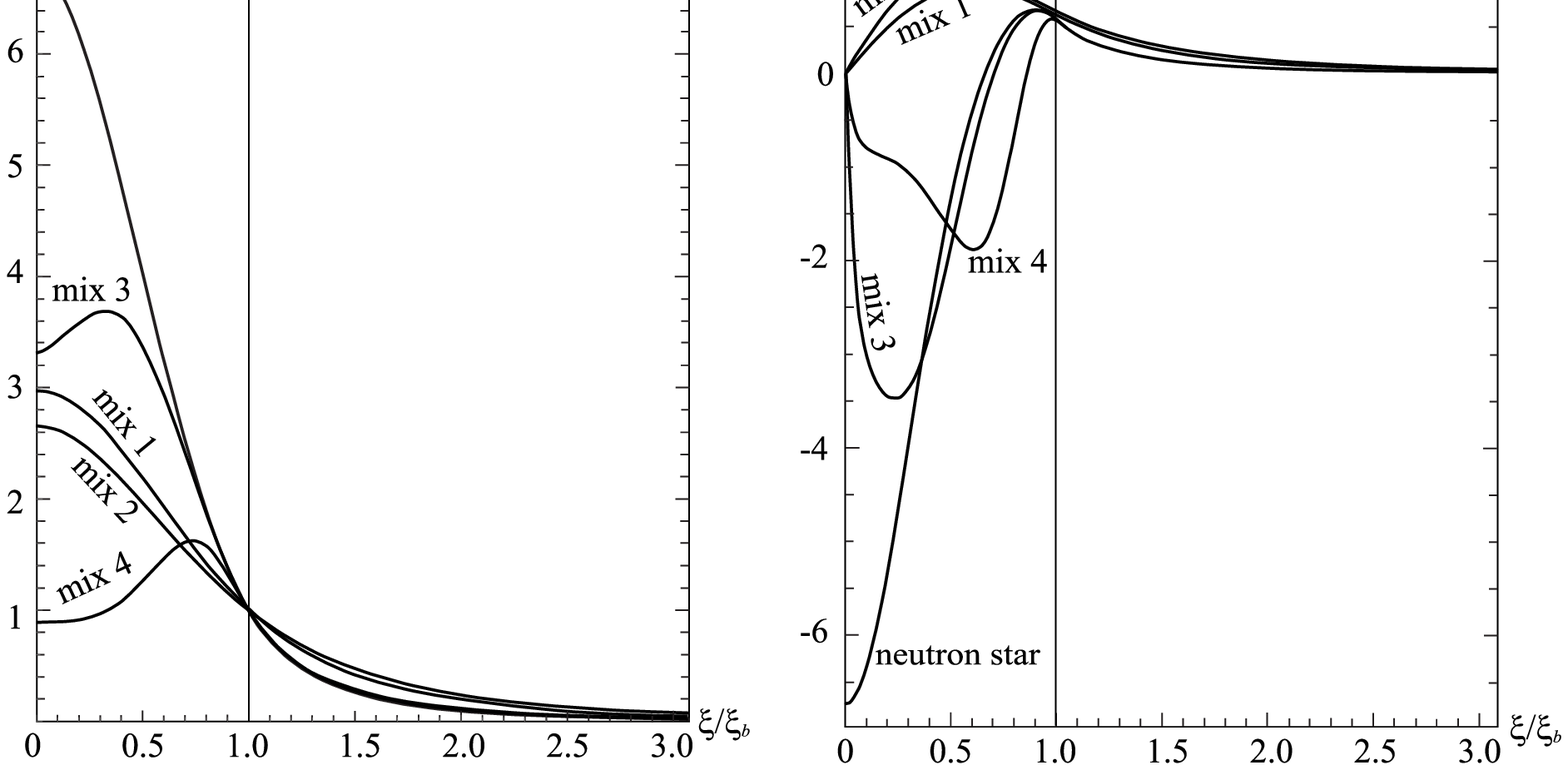}
\vspace{1.0cm}
%    \label{energ_n_1_5_fig}
  \end{center}
\end{minipage}\hfill
\begin{minipage}{1\linewidth}
  \begin{center}
  \includegraphics[width=8.5cm]{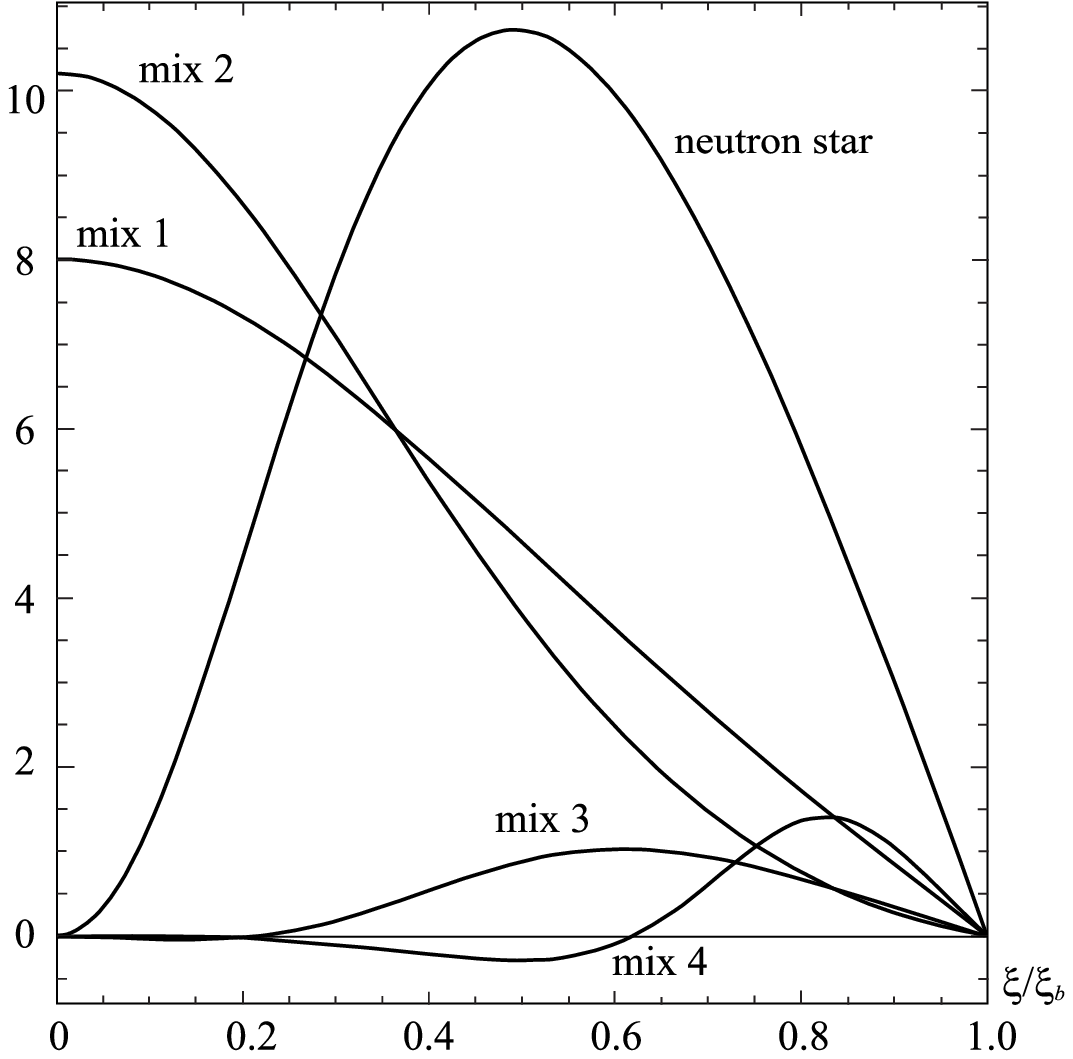}
\vspace{1.0cm}
  \end{center}
\end{minipage}\hfill
  \caption{The tetrad component $B_{\hat{r}}$ and $B_{\hat{\Theta}}$
of the magnetic field, evaluated
on the symmetry axis ($\Theta=0$) and
in the equatorial plane ($\Theta=\pi/2$), respectively,
(in units of the surface strength
of the magnetic field $B_s=10^{12}~\text{gauss}$),
and the dimensionless current $\bar{j}$
are shown versus the radial coordinate $\xi$ (in units of
the boundary value of the neutron fluid, $\xi_b$).
Configurations with the following parameters are exhibited:
for the neutron star we assume a central mass-density of the neutron matter
$\rho_{b c}= 5.37 \times 10^{14} \text{g cm}^{-3}$,
a length scale $L=10\, \text{km}$, and $\beta=0$ (isotropic fluid).
For the mixed systems we take:
for mix~1 -- $\rho_{b c}= 2.75 \times 10^{14} \text{g cm}^{-3}$, $L=10\, \text{km}$, and $\beta=0$;
for mix~2 -- $\rho_{b c}= 3.25 \times 10^{14} \text{g cm}^{-3}$, $L=10\, \text{km}$, and $\beta=0.4$;
for mix~3 -- $\rho_{b c}= 2.75 \times 10^{14} \text{g cm}^{-3}$, $L=1\, \text{km}$, and $\beta=0.4$;
for mix~4 -- $\rho_{b c}= 0.0258 \times 10^{14} \text{g cm}^{-3}$, $L=1\, \text{km}$, and $\beta=2$.
The systems mix~3 and mix~4 have a shifted maximum
of the neutron matter density.
All configurations have approximately the same total mass,
$M \approx 3.17 M_\odot$.
}
\label{MF_fig}
\end{figure}

\begin{figure}[h!t]
\centering
  \includegraphics[height=12.2cm]{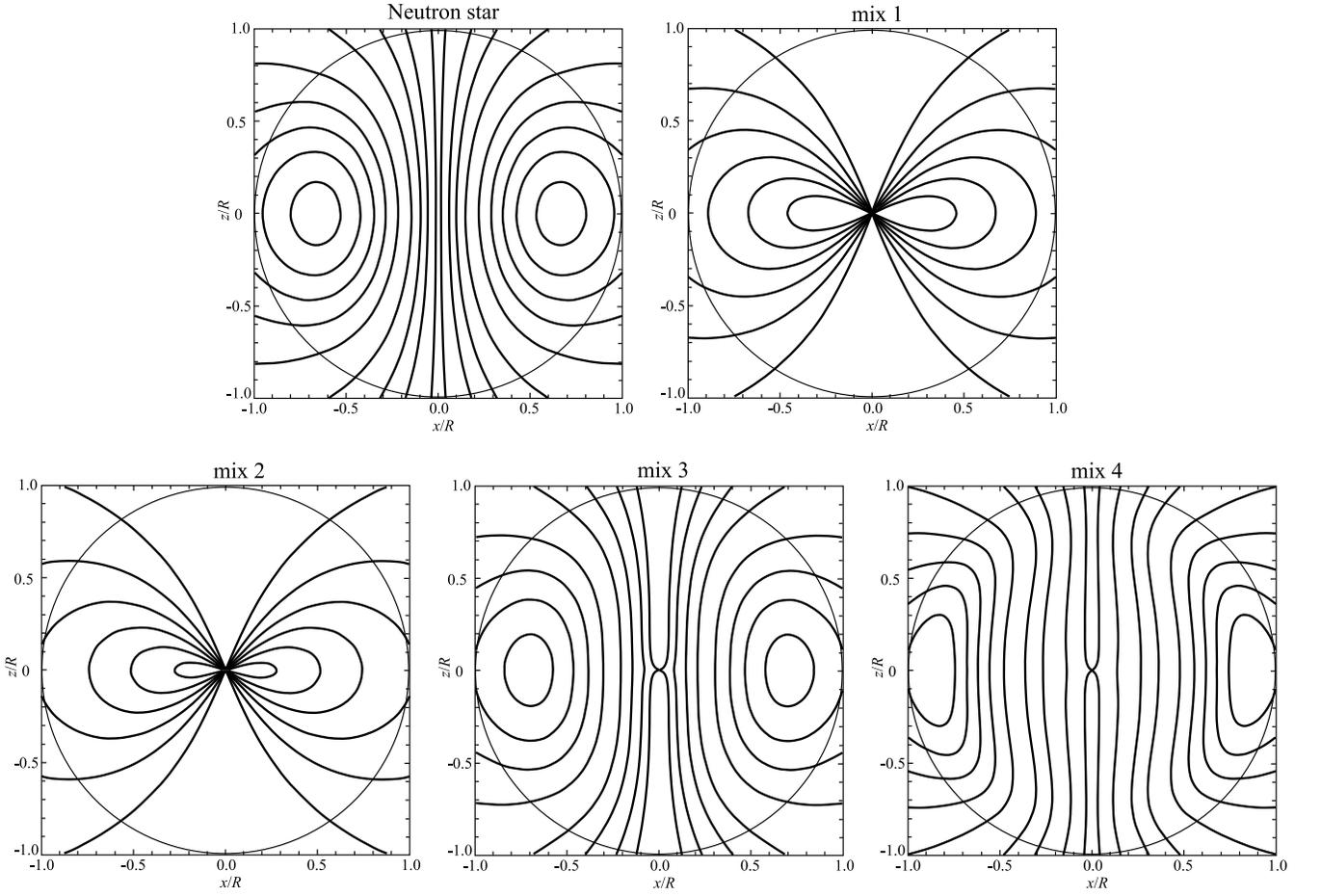}
%\vspace{0.5cm}
\caption{Magnetic field lines in the $x-z$ plane
are shown for the configurations of Fig.~\ref{MF_fig}.
The circles denote the boundary of the neutron fluid,
possessing radius $R$.
}
\label{fig_magn_field_lines}
\end{figure}

\begin{figure}[h!t]
\centering
  \includegraphics[height=7cm]{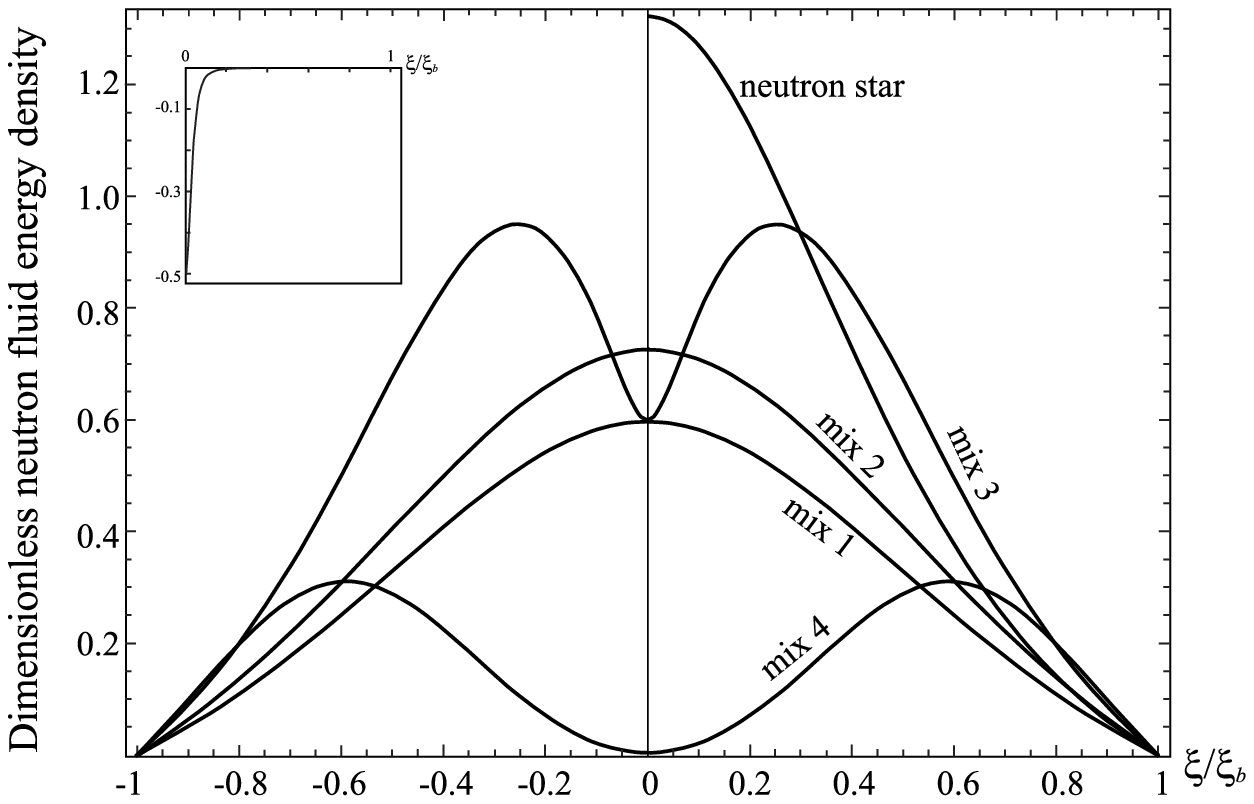}
%\vspace{-1.cm}
\caption{Energy density $T_{t (\text{fl})}^t$ of the neutron fluid
(in units of $\rho_{b c} c^2$ for the neutron star
and $\varphi_1^2$ for the mixed systems)
for the configurations of Fig.~\ref{MF_fig}.
For the systems mix~3 and mix~4
the energy density has been multiplied by a factor of 100
to make the scales of the graphs comparable.
The energy density of the scalar field (in units of $\varphi_1^2$)
is shown in the inset for the system mix~4.
}
\label{fig_NF_energ_dens}
\end{figure}

We now discuss the numerical results for the mixed systems
for isotropic and anisotropic neutron fluids.
We here consider only configurations with a single throat.
In particular, we illustrate the magnetic field
and the current for 4 systems, denoted mix~1 -- mix~4,
all possessing
approximately the same total mass $M \approx 3.17 M_\odot$.
While this mass is somewhat on the large side for
ordinary neutron stars, the properties of the
magnetic field obtained for these compact
objects appear to be generic.
Moreover,
all configurations are chosen such as to possess
approximately the same surface magnetic field
$B_s \approx 10^{12}~\text{gauss}$.

Consequently, a distant observer
would measure the same mass and surface magnetic field
for each of the four mixed configurations
as well as for the ordinary neutron star,
exhibited in Figs.~\ref{MF_fig}-\ref{fig_NF_energ_dens}.
While similar in these respects,
the neutron star and the mixed systems will differ,
however, strongly in other respects, as for instance
in their sizes.

%\subsubsection{Isotropic fluid ($\beta=0$)}
\begin{itemize}
\item Isotropic fluid ($\beta=0$)

For the first mixed system, mix~1,
we follow Ref.~\cite{Dzhunushaliev:2014mza},
and take a characteristic length $L=10\, \text{km}$.
To obtain a total mass of $M \approx 3.17 M_\odot$,
we choose $\rho_{b c}= 2.75 \times 10^{14} \text{g cm}^{-3}$.
The free parameters $a_c$ and $A_1$ associated with
the magnetic field are chosen such as to
provide
a surface magnetic field of strength $B_s \approx 10^{12}\text{gauss}$.
%for the mixed configuration.
(In particular, the surface value of the magnetic field strength
on the symmetry axis ($\Theta=0$) is chosen to
have this value.) %be $B_s=10^{12}~\text{Gauss}$.
%We do not assume any anisotropy.
The results of the calculations are shown
in Figs.~\ref{MF_fig}--\ref{fig_NF_energ_dens}
by the curves labeled by mix~1.

In Fig.~\ref{MF_fig} the value of the component
$B_{\hat{r}}$ on the symmetry axis ($\Theta=0$),
the value of the component $B_{\hat{\Theta}}$
in the equatorial plane ($\Theta=\pi/2$),
and the current $\bar j$ are shown.
While the component $B_{\hat{r}}$ is monotonic for the neutron star
as well as for the mixed configuration,
it is much smaller in the interior for the system mix~1.
The component $B_{\hat{\Theta}}$, on the other hand,
is equal to zero at the center of the mixed configuration.
This is in contrast to its large absolute value at
the center of an ordinary neutron star.
For the mixed configuration
the vanishing of $B_{\hat{\Theta}}$ at the center
results from the boundary conditions
\eqref{bound_cond_all} and \eqref{bound_cond_all2},
necessary for asymptotically flat solutions.

The currents $\bar j$ differ also considerably
for the mixed configuration and an ordinary neutron star,
as seen in Fig.~\ref{MF_fig}.
The current of the neutron star has its maximum
roughly in the middle of the star, $\xi \approx \xi_b/2$,
tending to zero at the center and at the boundary.
In contrast, for the mixed system mix~1 the current
rises monotonically, reaching its maximum at the center,
$\xi=0$.

The magnetic field lines are exhibited in Fig.~\ref{fig_magn_field_lines},
where the spatial coordinates are given in units of the radius
of the fluid, $R$.
The radius of the mixed system mix~1 is given by
$R \approx 16\, \text{km}$ as compared to
$R \approx 26\, \text{km}$ for the neutron star.
But apart from the different sizes, there is a profound
distinction in the field lines, since in the
mixed system all field lines penetrate the center,
i.e., the wormhole throat.

In Fig.~\ref{fig_NF_energ_dens} we exhibit the
energy density $T_{t (\text{fl})}^t$ of the neutron fluid.
The energy density is monotonic for both the
ordinary neutron star and the mixed system.
We note that the energy densities are given
in different units here,
$\rho_{b c} c^2$ for the neutron star
and $\varphi_1^2$ for the mixed system.

%\subsubsection{Anisotropic fluid ($\beta\neq 0$)}
\item Anisotropic fluid ($\beta\neq 0$)

As discussed above, in the presence of an anisotropy
there are two different types of solutions.
For ${\cal B}>{\cal B}_{p}$ the density of the neutron fluid
has its maximum $\rho_{\text{max}}$ at the center.
For ${\cal B}<{\cal B}_{p}$, on the other hand,
the maximum of the neutron fluid density
%$\rho_{\text{max}}$
is located away from the center.

\begin{itemize}
\item[(i)] ${\cal B}>{\cal B}_{p}$:\\
To obtain a total mass of $M \approx 3.17 M_\odot$,
we may choose, for example, the anisotropy parameter $\beta=0.4$
together with the central rest-mass density
$\rho_{b c}= 3.25 \times 10^{14} \text{g cm}^{-3}$.
We refer to the resulting system as mix~2.
The curves labeled by mix~2 in Fig.~\ref{MF_fig}
show the magnetic field components and the current
for this mixed system.
In spite of the anisotropy,
the magnetic field and the current are similar to
those of the isotropic case, i.e., mix~1.

This holds also for the magnetic field lines
shown in Fig.~\ref{fig_magn_field_lines},
where the neutron fluid has a radius
$R \approx 13\, \text{km}$.
Likewise, the energy density $T_{0 (\text{fl})}^0$ of the neutron fluid
shown in Fig.~\ref{fig_NF_energ_dens}
is similar for mix~1 and mix~2.
Typical distributions of the total energy density
(with the energy density of the scalar field also taken into account)
can be found in Ref.~\cite{Dzhunushaliev:2014mza}.

\item[(ii)]
${\cal B}<{\cal B}_{p}$:\\
One can obtain configurations with a shifted maximum,
resulting from a sufficiently small value of the parameter ${\cal B}$,
for example,
by reducing the value of the characteristic length $L$,
e.g., for the choice $L=1\, \text{km}$,
selected here.
In order to obtain configurations
with a total mass of $M \approx 3.17 M_\odot$,
we here consider two values of the anisotropy parameter $\beta$:
$\beta=0.4$ with $\rho_{b c}= 2.75 \times 10^{14} \text{g cm}^{-3}$ and
$\beta=2$ with $\rho_{b c}= 0.0258 \times 10^{14} \text{g cm}^{-3}$.
The resulting mixed configurations
are labeled by mix~3 and mix~4 in the figures.

Fig.~\ref{MF_fig} shows that
$B_{\hat{r}}$ now assumes its maximum also away from the center
of the configuration. For mix~3 its maximum is roughly at
$\xi_b/2$, while for mix~4 it is still further outwards.
Likewise, $B_{\hat{\Theta}}$ assumes a minimum
away from the center, roughly at
$\xi_b/4$ for mix~3, while again further outwards for mix~4.

Also the current is no longer maximal at the center,
%as for mix~1 and mix~2,
but it assumes its maximum values
close to those of $B_{\hat{r}}$ for mix~3 and mix~4.
Moreover, it is interesting to note
that for the systems mix~3 and mix~4
the current can take negative values in the inner regions.
Getting close to zero at the center,
the current of these configurations
is thus more akin to the current of a neutron star
than to the current of the mixed configurations
mix~1 and mix~2, where the current becomes
maximum at the center.

This is reflected in Fig.~\ref{fig_magn_field_lines},
where the magnetic field lines of
the configurations mix~3 and mix~4 resemble more those
of the neutron star. Thus only little magnetic
flux penetrates the throat.
The neutron fluids of these configurations have radii
$R \approx 24\, \text{km}$ for mix~3 and
$R \approx 27\, \text{km}$ for mix~4.
Finally, the energy density $T_{t (\text{fl})}^t$ of the neutron fluid
shown in Fig.~\ref{fig_NF_energ_dens}
exhibits a local minimum at the center, i.e., at the throat,
while its maximum is shifted away from the center,
as required.

Finally we note that for the systems mix~3 and mix~4
the energy density of the neutron fluid in the central region
is small as compared to the energy density of the scalar field,
shown in the inset in Fig.~\ref{fig_NF_energ_dens}.
For larger values of $\xi$, however, it becomes dominant,
providing most of the total mass of the system.

\end{itemize}
\end{itemize}

\section{Stability} % of the system with shifted maximum of the neutron fluid density}
\label{stabi}

Let us now address the stability of the mixed configurations,
focussing on the aspect of anisotropy,
and compare with the stability of ordinary neutron stars.
We here do not consider the influence of the magnetic field,
since its contribution to the energy balance of these
configurations is negligible, as long as we consider values of the
surface field strength on the order of $B_s=10^{12}~\text{gauss}$.
Consequently, also the deformation of the configurations
is negligible, and we address in the following only the stability
of the spherically symmetric background configurations.

\subsection{Neutron stars}

For a given EOS, neutrons stars form a family of configurations,
characterized by the value of their central density.
As the central density of ordinary neutron stars increases,
their mass first increases, reaches a maximum value $M_{\rm max}$,
and then decreases again.
The maximum value of the mass is crucial from the stability point of view,
since at $M_{\rm max}$
%which is associated with a certain value of the central density,
the stability of the neutron stars changes.
Denoting the central density associated with $M_{\rm max}$
as the critical density,
a linear stability analysis reveals that at the critical density
the eigenvalue of the radial mode changes sign, making neutron stars
with a larger central density unstable.

Neutron stars with a shifted maximum of the neutron fluid density,
as caused by anisotropy, were investigated in Ref.~\cite{Hillebrandt:1976}.
Here the dependence of the mass on the maximum value of the
fluid density was considered
(instead of the central value as in the isotropic case).
By studying linear radial oscillations of these stars,
the authors of~\cite{Hillebrandt:1976} showed
that the square of the lowest eigenfrequency of these oscillations
changes sign at the maximum value of the mass $M_{\rm max}$,
completely analogous to the isotropic case.
Thus the well-known criterion for stability,
according to which configurations with a central density
below (above) a critical value are stable (unstable),
generalizes to the anisotropic case,
when the {\sl central density} is replaced by
the {\sl maximum density}.

\subsection{Mixed configurations}

For the mixed configurations
a linear stability analysis has been performed in
Ref.~\cite{Dzhunushaliev:2014mza}
for isotropic and anisotropic neutron fluids.
In the latter case, however, the maximum value
of the density was not shifted away from the center.
Thus only families of configurations of the type
mix~1 and mix~2 were considered.
Here the square of the lowest eigenfrequency
of the radial modes was calculated
and all of these configurations were shown to be unstable.
In fact, the unstable mode of the wormhole
\cite{Gonzalez:2008wd,Bronnikov:2011if}
is inherited by the mixed neutron-star-plus-wormhole
configurations~\cite{Dzhunushaliev:2014mza}.

Let us now consider the stability of mixed configurations
with an anisotropic neutron fluid and a maximum fluid density
shifted away from the center, i.e., configurations
of the type mix~3 or mix~4.
Performing a mode analysis of these systems reveals, however,
that the location of the maximum fluid density is not relevant
for the stability.
The eigenvalue of the set of equations is always negative.
Thus the wormhole with its negative mode dominates the
stability properties. The negative mode is inherited,
independent of the particular properties of the fluid.

\begin{table}[h]
%{\footnotesize
\caption{Characteristics of configurations
of the type mix~4 with $\beta=2$ with a shifted maximum
of the density of the neutron fluid (see Fig.~\ref{fig_NF_energ_dens}).
Here the central rest-mass density $\rho_{b c}$,
the maximum density $\rho_{\text{max}}$ of the neutron fluid
(both in units of $10^{14} \text{g cm}^{-3}$),
the total mass $M$ (in solar mass units),
the radius of the neutron fluid $R$ (in kilometers),
and the square of the lowest eigenfrequency $\omega_0^2$ are shown.}
\vspace{.3cm}
\begin{tabular}{p{1.5cm}p{1.5cm}p{1.5cm}p{1.5cm}p{1.5cm}}
\hline \\[-5pt]
$\rho_{b c}$ & $\rho_{\text{max}}$ &  $ M/M_\odot$ & $ R$   & $\omega_0^2$ \\[2pt]
\hline \\[-7pt]
0.01&0.713&	2.205&	29.992 & -0.45\\
0.0258 &1.529& 3.172&	27.105 & -0.32\\
0.06&2.772&	3.572&	23.876 & -0.20\\
0.09&3.588&	3.582&	22.245 & -0.16 \\
0.15&4.855&	3.462&	20.218 & -0.12\\
0.30&7.062&	3.150&	17.631 & -0.082\\
2.75&18.837& 1.937&	11.485 & -0.028\\
\hline
\end{tabular}
\label{tab1}
%}
\end{table}

We illustrate the instability of this type of configurations
in Table~\ref{tab1},
where we show a set of relevant properties for solutions
associated with the mix~4 parameter set,
i.e., for a family of configurations with length scale $L=1\, \text{km}$
and anisotropy parameter $\beta=2$.
In particular, we exhibit the
central rest-mass density $\rho_{b c}$,
the maximum density $\rho_{\text{max}}$,
the mass $M$, the radius $R$ and the eigenvalue
of the stability equations, i.e., the square
of the lowest eigenfrequency $\omega_0^2$.

As seen in the table, while both densities increase
the total mass of the system approaches a maximum $M_{\rm max}$
at a critical value of
$\rho_{b c}$ with an associated $\rho_{\text{max}}$.
However, the eigenvalues are always negative,
and thus the systems under consideration are all linearly unstable.
Consequently, the criterion for stability
based on the dependence of the mass $M$
on the rest-mass density
(central $\rho_{b c}$ or shifted $\rho_{\text{max}}$)
cannot be applied for mixed neutron-star-plus-wormhole systems.

\section{Conclusion}
\label{conclusion}

In the present paper we studied equilibrium
magnetized mixed configurations
consisting of a wormhole threaded by a neutron fluid
with an isotropic or anisotropic pressure.
The fluid is described by the simplest relativistic
polytropic EOS \eqref{eqs_NS_WH}.
In contrast to ordinary neutron stars, such
mixed configurations possess a nontrivial topology of the spacetime
permitted by the presence of exotic matter
in the form of a massless ghost scalar field.

Our goal was to investigate the influence
of the nontrivial topology
on the structure of the magnetic field,
which was modeled here in the form of
an axisymmetric, poloidal magnetic field
produced by a toroidal electric current.
For the sake of comparison,
we considered also an ordinary magnetized neutron star
(with trivial topology),
described by the same EOS, with the same
form of the magnetic field,
created by the same type of electric current.

For both types of configurations, the ordinary neutron star
and the mixed systems,
the physical parameters were chosen in such a way
that they would look similar from the point of view
of a distant observer,
i.e., they should possess the same masses
and surface strengths of the magnetic field.

The results can be summarized as follows:
\begin{enumerate}
\itemsep=-0.2pt
\item[(i)] %1.
In the case of an isotropic fluid,
there are substantial differences in
the magnetic field and the current
of a mixed system as
compared to an ordinary neutron star:
the mixed system has a considerably smaller central value of
the radial component $B_{\hat{r}}$, as seen in Fig.~\ref{MF_fig}.
At the same time, the component $B_{\hat{\Theta}}$
tends to zero at the center,
in contrast to an ordinary neutron star,
where $|B_{\hat{\Theta}}|$ becomes maximum at the center.
The magnetic field lines of this system differ strongly
from those of ordinary neutron stars,
with all of the magnetic flux penetrating the throat.

\item[(ii)]
In the case of an anisotropic neutron fluid,
there are two types of solutions,
where the neutron fluid has a maximum density
either at the center of the system or
shifted away from the center, as seen in Fig.~\ref{fig_NF_energ_dens}.
In the first case,
the magnetic field and the current are very similar to those
of a mixed system with an isotropic fluid,~(i).
%, but these systems have substantially smaller masses.
In the second case, however,
the magnetic field and the current differ quite strongly
from those obtained for %isotropic mixed systems,
systems whose maximum density is residing at the center,
as seen in Fig.~\ref{MF_fig}.
For systems with a shifted maximum density
the current can take negative values
in the interior region, and very small values at the center.
%as in the case of ordinary neutron stars.
The magnetic field lines of these systems then are much
more akin to those of ordinary neutron stars,
with only little magnetic flux penetrating the throat.

\item[(iii)]
A linear stability analysis of the mixed systems
revealed their instability,
independently of the location of the
maximum of the neutron fluid density.
This pervasive instability is in contrast to ordinary neutron stars,
which always possess a stable branch,
also in the case of anisotropic neutron fluids~\cite{Hillebrandt:1976}.
Clearly, the instability of the wormhole
is inherited by the mixed neutron-star-plus-wormhole systems,
and the various properties of the neutron matter cannot
overturn this instability.
\end{enumerate}

Thus the studies showed that the structure
of the interior magnetic field may depend substantially
not only on the form of the electric currents in the interior
and the physical properties of the neutron matter
(given by the choice of the equation of state,
as demonstrated, for example,
in Refs.~\cite{Bocquet:1995je,Kiuchi:2007pa})
but also on the topology of the spacetime.
On the other hand, the parameters of the
mixed neutron-star-plus-wormhole systems can be selected
such that from the point of view of a distant observer
there will be no substantial external differences between ordinary
magnetized neutron stars and mixed systems with nontrivial topology.

\section*{Acknowledgements}

We gratefully acknowledge support provided by the Volkswagen Foundation
and by the DFG Research Training Group 1620 ``Models of Gravity''.
This work was partially supported by a grant within FP7, Marie Curie
Actions, People, International Research Staff Exchange
Scheme (IRSES-606096)
and by a grant in fundamental research in natural sciences
of the Ministry of Education and Science of Kazakhstan.

\end{document}